**Title**: Value-transforming financial, carbon and biodiversity footprint accounting

**Authors:** Sami El Geneidy[1,2], Maiju Peura[1,3], Viivi-Maija Aumanen[4], Stefan Baumeister[1,2], Ulla Helimo[1,3,4], Veera Vainio[1,3], Janne S. Kotiaho[1,3]

**Institutions:**

[1] School of Resource Wisdom, University of Jyväskylä, P.O. Box 35, FIN-40014 University of Jyväskylä, Finland

[2] School of Business and Economics, University of Jyväskylä, P.O. Box 35, FIN-40014 University of Jyväskylä, Finland

[3] Department of Biological and Environmental Science, University of Jyväskylä, P.O. Box 35, FIN-40014 University of Jyväskylä, Finland

[4] Division of Policy and Planning, University of Jyväskylä, P.O. Box 35, FIN-40014, Finland

**Corresponding Author:**

Sami El Geneidy

Seminaarinkatu 15, University of Jyväskylä, P.O. Box 35, FIN-40014, Finland

sami.s.elgeneidy@jyu.fi

ORCID: https://orcid.org/0000-0003-4408-5256

**Conflict of Interest Statement:** The authors declare no conflict of interest.

**Data Availability Statement:** The data that support the findings of this study are openly available in Zenodo at https://doi.org/10.5281/zenodo.8369650. The organizational data that support the findings of this study are available on request from the corresponding author. The data is not publicly available due to privacy or ethical restrictions.

**Abstract:** Transformative changes in our production and consumption habits are needed to halt biodiversity loss. Organizations are the way we humans have organized our everyday life, and much of our negative environmental impacts, also called carbon and biodiversity footprints, are caused by organizations. Here we explore how the activity data of any organization can be exploited to develop integrated carbon and biodiversity footprints. As a metric we utilize spatially explicit potential global loss of species across all ecosystem types and argue that it can be understood as the biodiversity equivalent. The utility of the biodiversity equivalent for biodiversity could be like what carbon dioxide equivalent is for climate. We provide a global country-specific dataset that organizations, experts and researchers can use to assess biodiversity footprints of human activities. We also argue that the current integration of financial and environmental accounting is superficial, and provide a framework for a more robust financial value-transforming accounting model. To test the methodologies, we utilized a Finnish university as a living lab. Assigning an offsetting cost to the footprints significantly altered the financial value of the organization. We believe such value-transforming accounting is needed to draw the attention of senior executives and investors to the negative environmental impacts of their organizations.

## 1. Introduction

Biodiversity loss is driven by human land and sea use and their changes, direct exploitation of nature, climate change, pollution, and introduction of invasive alien species (IPBES, 2019). These direct drivers result from various underlying root causes such as human population dynamics, consumption patterns, trade, and governance, which are in turn underpinned by societal values and behaviours (Díaz et al., 2019; IPBES, 2019; Visseren-Hamakers et al., 2021). Managing the direct drivers of biodiversity loss alone will not produce sustained outcomes sufficient to bend the curve of biodiversity loss (Leclère et al., 2020; Mace et al., 2018). Instead, we must direct our efforts to the root causes such as consumption and trade.

Everyday life and the economics of societies are organized through organizations, be they private businesses, public services, or non-governmental organizations. Only the direct emissions of around 9 000 companies contributed to over 38% of global greenhouse gas emissions in 2021 (CDP, 2022). In addition, the negative environmental impacts of nearly any organization extend through international trade and supply chains to all over the planet (Hong et al., 2022; Marques et al., 2019; Presberger & Bernauer, 2023).

Carbon and biodiversity footprint assessments are tools that can be used to investigate the negative environmental impacts of organizations. While carbon footprint assessments are abundant (Chen et al., 2021; Peters, 2010; Shi & Yin, 2021) and a few biodiversity footprint assessments have been attempted (Bull et al., 2022; Pokkinen et al., 2024; Taylor et al., 2023), there is a lack of universal approaches suitable for all kinds of organizations especially in the global context. Furthermore, in a globalized economy the comparability of the biodiversity footprints of different organizations and their value chains in different regions of the world remains difficult (Bromwich et al., 2025; Sanyé-Mengual et al., 2023), hindering the accountability of organizations to policymakers and the public.

Environmental accounting, such as carbon and biodiversity footprinting, seems to remain isolated within organizations and even when it is integrated with other reporting practices like financial reports it can still remain unexploited in management decisions (Bracci & Maran, 2013; Maas et al., 2016; Saravanamuthu, 2004; Veldman & Jansson, 2020). Indeed, decision-making in organizations is ultimately guided by information obtained from financial accounts (Bracci & Maran, 2013; Hines, 1988; Saravanamuthu, 2004; Schaltegger & Burritt, 2000; Veldman & Jansson, 2020). Thus, merely focusing on environmental accounting is unlikely to steer organizations towards the transformative changes necessary to reach a Nature Positive future (Booth et al., 2024). A shift is needed in how we do and view both environmental and financial accounting. An integration between environmental and financial accounting may be achieved when the biodiversity and carbon footprints have a cost that becomes visible in the income statement, a key output in the financial accounting system of an organization.

### 1.1 From regional to global indicators of biodiversity loss: the biodiversity equivalent

Research on biodiversity footprints has emerged in recent years, resulting in a variety of biodiversity footprinting methods (e.g. Crenna et al., 2020; Damiani et al., 2023; Marques et al., 2017). The most extensive life cycle impact assessment (LCIA) methods, such as ReCiPe (Huijbregts et al., 2017), LC-IMPACT (Verones et al., 2020), and Impact World+ (Bulle et al., 2019) cover multiple drivers of biodiversity loss and consider spatially explicit impacts on terrestrial, freshwater, and marine ecosystems, encompassing a wide range of taxa (Damiani et al., 2023). Differences between these methods exist, particularly in how they model the biodiversity impacts caused by the drivers of biodiversity loss (Bromwich et al., 2025; Marquardt et al., 2019; Sanyé-Mengual et al., 2023).

The Potentially Disappeared Fraction of Species (PDF) is a commonly used metric for biodiversity footprinting (Crenna et al., 2020; Marques et al., 2017). It can model the share of species at risk of extinction regionally (Bulle et al., 2019; de Baan et al., 2013; Huijbregts et al., 2017) or globally (Verones et al., 2020) due to specific drivers of biodiversity loss, such as land use. Another commonly used metric is the Mean Species Abundance (MSA) (Alkemade et al., 2009; Schipper et al., 2020). MSA measures the average abundance of species relative to a reference state on a regional scale (Alkemade et al., 2009; Schipper et al., 2020). The basic building blocks of both regional and global biodiversity footprint indicators are presented in Figure 1.

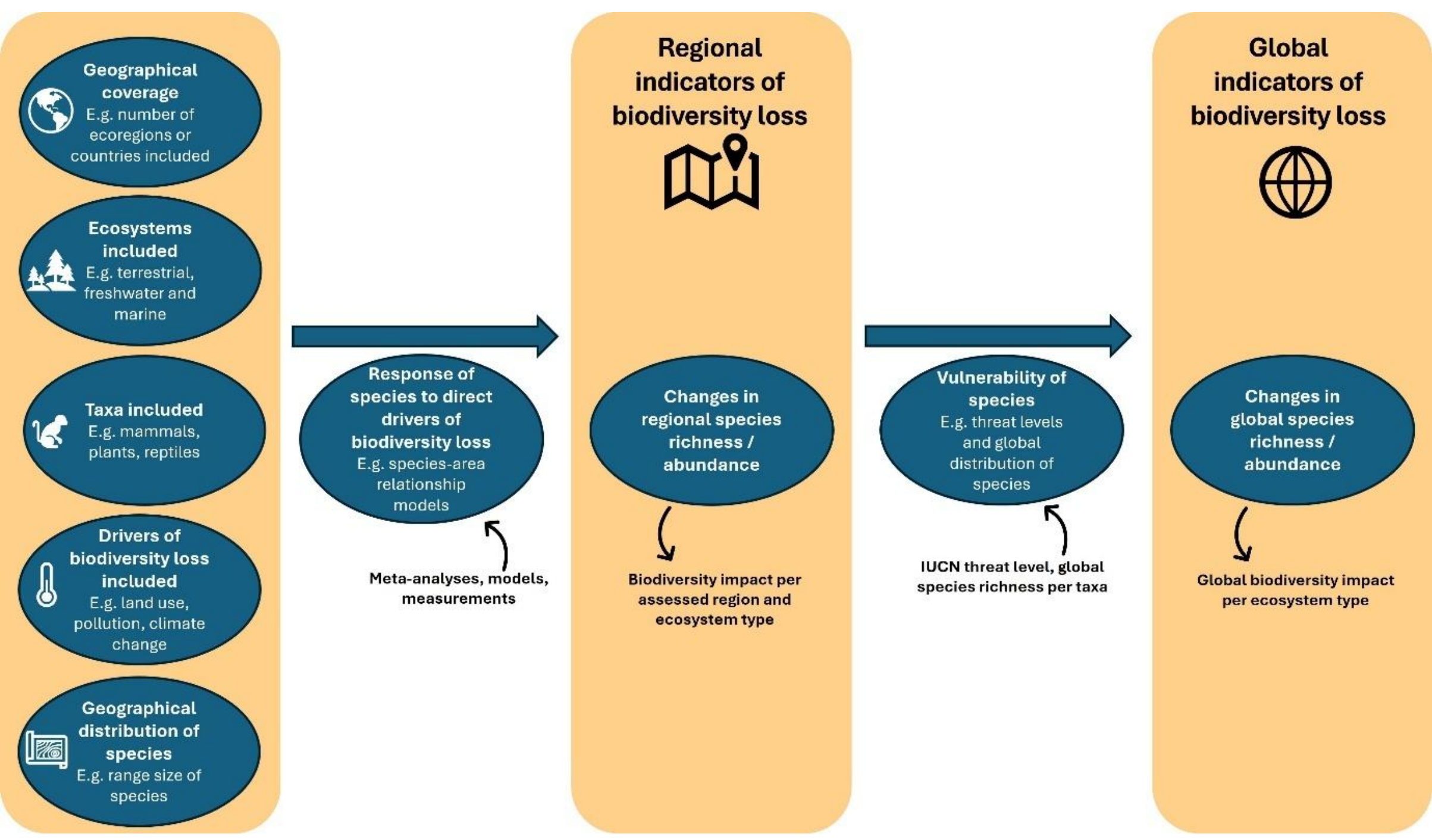

**Fig. 1** The common components of biodiversity footprint indicators and the relationship between regional and global indicators of biodiversity loss

We provide an example to further illustrate the differences between regional and global indicators of biodiversity loss. Consider an organization that has a land use impact in two regions of the world, for example Finland and Brazil. If the condition of the impacted area in Finland is assumed to be 0.25 MSA (25 % of species abundance remains before the impact compared to the reference condition and are completely lost due to the impact) and the impacted area is 200 hectares (ha), the biodiversity footprint of the organization would be 50 MSA×ha. If the condition of the impacted area in Brazil would then be 0.50 MSA (50 % of species abundance remains before the impact compared to the reference condition and are completely lost due to the impact) and the impacted area would be 100 ha, the biodiversity footprint of the organization in Brazil would be 50 MSA×ha. Thus, with the regional MSA indicator the biodiversity footprint of the organization in Finland and in Brazil appear to be of the same magnitude, considering, however, that the numbers above are merely example values to illustrate the functionality of the indicator. However, the magnitudes of the biodiversity footprints differ from each other if the impacts are considered from the global species' perspective. Using the LC-IMPACT database and the global PDF indicator (Verones et al., 2020) with the above example numbers the organization's biodiversity footprint due to land use in Finland would be 5.30E-11 global PDF (0.000000000530 % of global terrestrial species would be lost), while the biodiversity footprint in Brazil would be 2.24E-09 global PDF (0.0000000224 % of global terrestrial species would be lost). The global biodiversity footprint in Brazil would be around 42 times larger than the footprint in Finland.

The fundamental difference between regional and global indicators of biodiversity loss is that they look at biodiversity loss from a different perspective. Regional indicators do not consider the global distribution or vulnerability of species, thus giving equal weight to different regions of the world. On the other hand, the global indicators weigh different regions of the world with different significance, based on the distribution and vulnerability of species. The use of regional and global metrics may pose significant implications for decision-making in organizations. For example, if two organizations assess the biodiversity footprint of their value chain using a regional metric, such as the regional MSA, they may get the same biodiversity footprint value even though the impacts would be far from the same from a global perspective. In other words, for an organization with global value chains, it would be difficult to prioritize where actions are needed to best tackle global biodiversity loss. Furthermore, to effectively set organizational targets with regional indicators, the organization would need to separately address each region in its value chain, which could be inconvenient for efficient strategic decision-making. Even though the numbers naturally differ case by case, our simple example illustrates the significantly different interpretations regional and global indicators might give for similar activities across the globe. In a globalized economy, negative biodiversity impacts caused by production and consumption are distributed globally due to international trade flows and value chains (Koslowski et al., 2020; Lenzen et al., 2012; Marquardt et al., 2021; Wilting et al., 2017). There is a clear need for biodiversity footprint methods and indicators that can be used to support decision-making in and between organizations, especially to mitigate, compare and track biodiversity impacts across global value chains.

An issue that has been present in both regional and global indicators of biodiversity loss is the distinction between different ecosystem types, namely terrestrial, freshwater and marine. This produces complexity for decision-making as biodiversity footprint results are generally reported as three different values (Verones et al., 2020). We suggest that by developing the global PDF indicator further by weighting the ecosystem-specific biodiversity footprints with the number of species estimated to exist in terrestrial, freshwater and marine ecosystems (Román-Palacios et al., 2022), we can devise a globally unified, yet spatially explicit, indicator called the biodiversity equivalent (Figure 2). In essence, the biodiversity equivalent tells what fraction of the species of the world are at risk of going extinct globally due to the consumption and other activities of humanity. Since regionally lost species can still be recovered while globally extinct species are permanently lost, it has been argued that we should strive to devise and use indicators that estimate global species extinction risks (Verones et al., 2020), or indicators that translate regional species extinction to potential global extinction probabilities (Kuipers et al., 2019; Verones et al., 2022).

An important characteristic of the biodiversity equivalent is that it is influenced by the uneven distribution of species on the planet. Biodiversity equivalent is thus able to capture the fundamental understanding that the same human pressure in different regions of the world has a different potential to harm as well as conserve global biodiversity (Harfoot et al., 2021).

For example, if one hectare of land is transformed for intensive forestry in any given country, the same area transformed causes less global biodiversity loss in relatively species poor areas than it causes in relatively species rich areas. On the other hand, if both areas experienced a loss of the same amount of biodiversity equivalent, this would indicate that both areas experienced the same amount of global biodiversity loss. Different species would be lost in different parts of the world, but the fraction of globally potentially lost species would be the same. Consequently, we claim that as an indicator, the biodiversity equivalent has desirable characteristics much like carbon dioxide equivalent in that it provides a unified way of measuring biodiversity loss across the planet. We further justify the analogy between carbon dioxide equivalent and biodiversity equivalent in Supplementary Information SI1.

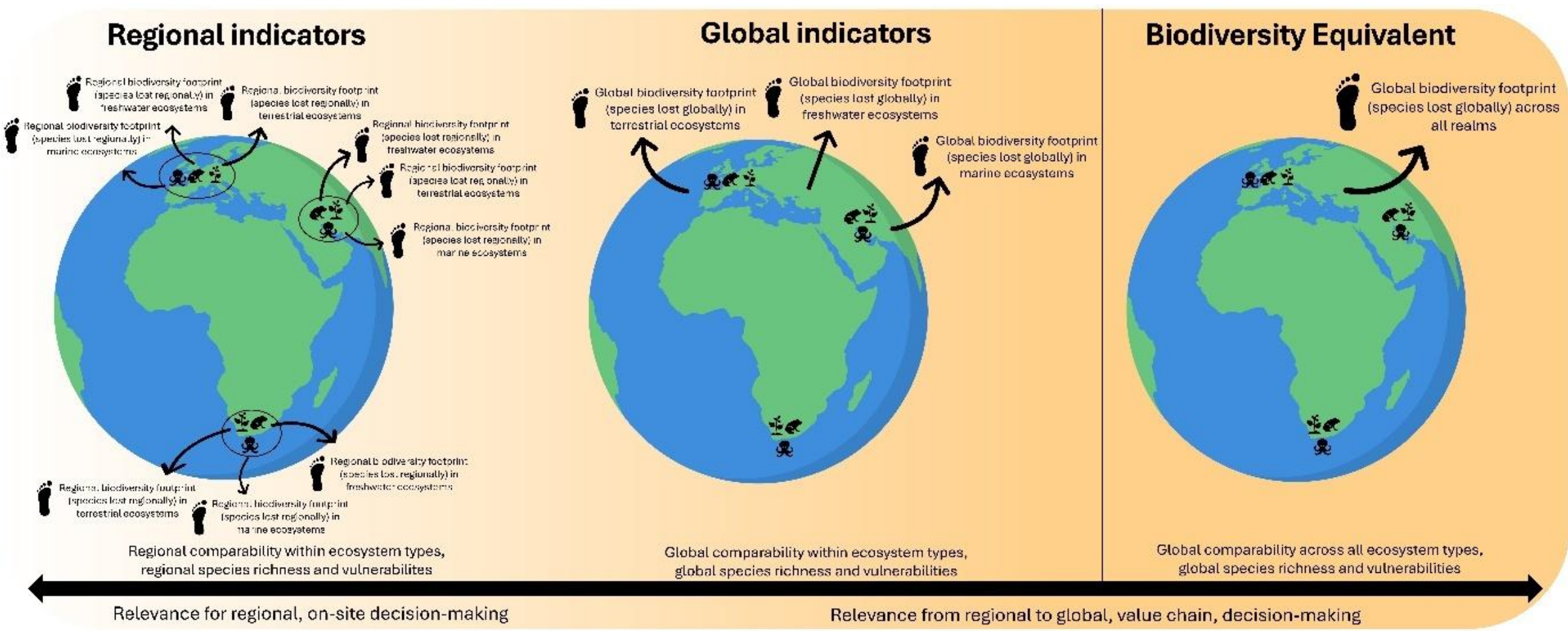

**Fig. 2** An illustration of the relationship between regional and global indicators of biodiversity loss and the biodiversity equivalent

Since the biodiversity equivalent combines biodiversity impacts across all regions and all ecosystem types under the same indicator, actions to mitigate the biodiversity footprint can be communicated efficiently to top management, policymakers, and citizens. The proposed metric addresses the international and multidimensional nature of biodiversity, giving organizations and policymakers a much-needed approach to identify, report and compare biodiversity impacts across global value chains. We provide a practical example of the utility of the biodiversity equivalent in an organization in Results.

## 2. Methods: Biodiversity Equivalent Impact Assessment

Next, we will present how the biodiversity footprint of an organization can be assessed in practice and how the biodiversity equivalent is calculated and reported to communicate results efficiently. Note that as climate change is one of the drivers of biodiversity loss, assessment of carbon footprints is an integral part of the methodology and will be calculated alongside the biodiversity footprints. We call the methodology the Biodiversity Equivalent Impact Assessment method, BIOVALENT. We will discuss how the results can be used to initiate value-transforming accounting practices in organizations.

Even though the biodiversity equivalent metric in the BIOVALENT method could also be used to assess local, on-site, biodiversity impacts, we believe its utility can be best explained by focusing on the value chain impacts (especially upstream), which is the focus of the case study presented in Results. The value chain perspective chosen here may be further justified by arguing that consumer choices and demand drive production processes (Lenzen et al. 2007, 2012). However, producers control production methods and thus, focus should ultimately be given to both production and consumption, i.e. the whole value chain (Lenzen et al., 2007).

Biodiversity footprint assessment and consequently the value-transforming integration of financial and environmental accounting in organizations requires essentially six steps, which are highlighted in Figure 3. The steps are meant to portray the biodiversity footprint assessment of an organization but can be applied to any kind of human activity. Furthermore, even though we present the whole process from the beginning to its end with enough technical detail allowing experts to reproduce the analysis, it is worth noting that the product of the steps 2 and 3, and to certain extent step 4, are reported in this paper in the form of the biodiversity impact factors (BIOVALENT database: https://oscsolutions.cc.jyu.fi/wisdom/biovalent/, all data available in the Zenodo repository: https://doi.org/10.5281/zenodo.8369650), so that organizations can use them directly to finalize the other steps of the process.

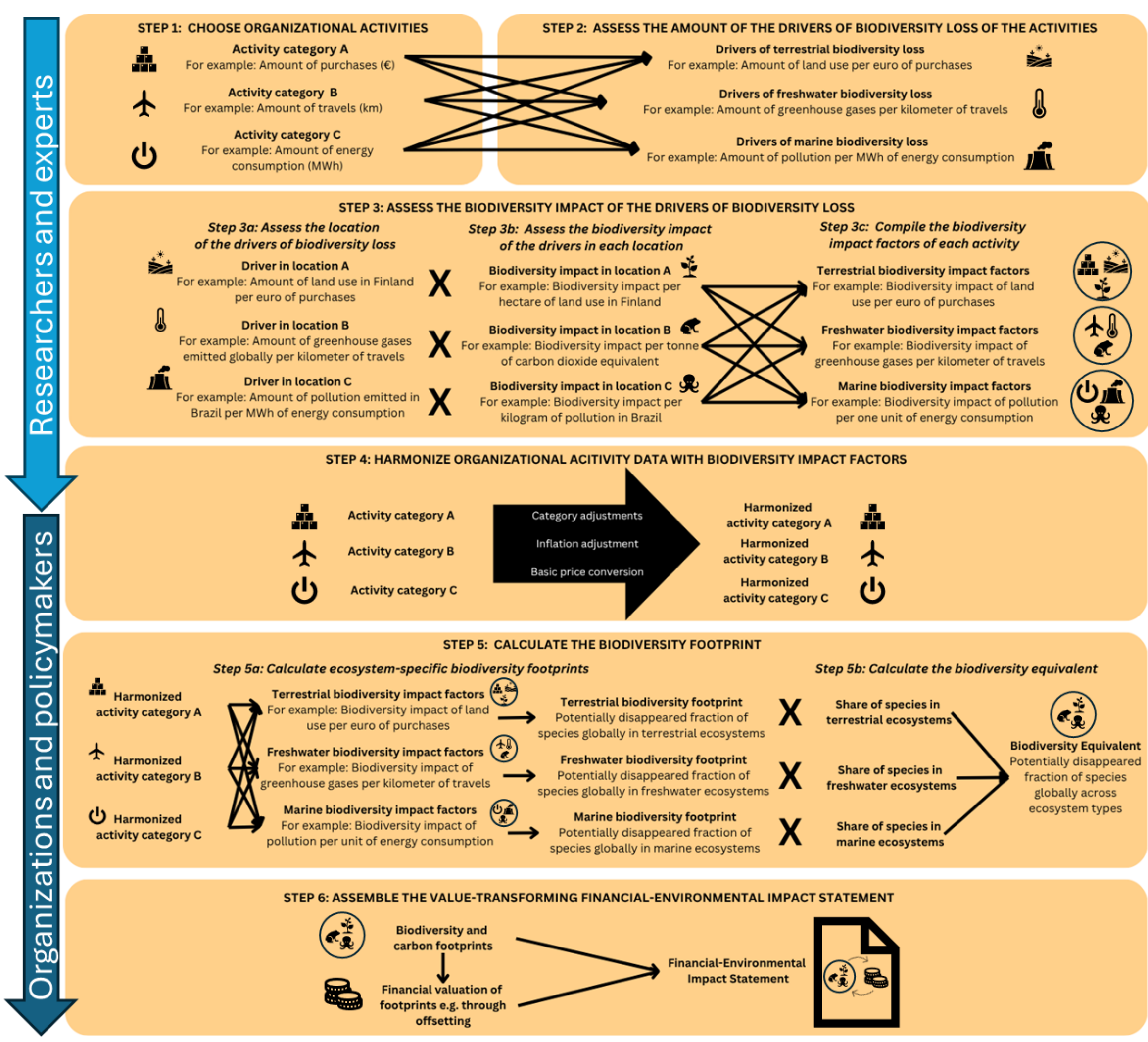

**Fig. 3** A schematic figure detailing the steps to assess the biodiversity footprint of an organization or any human activity with the BIOVALENT method and how to assemble a value-transforming financial-environmental impact statement

**STEP 1: Choose organizational activities to be assessed**

Organizations have various activities which are potentially measured and reported in different datasets and accounts. Financial accounts are universally the most dominant form of accounts. Other examples include datasets of energy consumption, water consumption, datasets of travelled kilometers or datasets of food consumption to name a few. The system boundaries of the assessment (i.e. what is included and excluded from the assessment) are essentially set by the financial accounts. Even though, for example distinct energy consumption datasets can be used to assess carbon and biodiversity footprints, energy consumption is part of financial accounts and can thus be identified as part of the system boundaries. Essentially this step is similar to goal and scope definition in life cycle assessment (Hellweg & Milà i Canals, 2014).

When choosing the organizational activities to be assessed, key considerations include, for example, data availability, accuracy, and extent. Trade-offs between accuracy and availability are common. For example, more detailed information about the types or material volumes of purchases is likely to provide more accurate results for environmental accounting than information provided by a financial account, but the compilation of such data can in many cases be exclusively work-intensive under the current data collection systems and is thus not often available. Nevertheless, when environmental impacts of an organization are assessed, it is essential to be able to account for all possible impacts, which might necessitate using a combination of primary data (e.g. energy consumption) and secondary (e.g. financial accounts) but more readily available data.

**STEP 2: Assess the amount of the drivers of biodiversity loss caused by the activities**

When the organizational activities and datasets have been identified, the drivers of biodiversity loss caused by the different types of activities have to be assessed. Essentially, one needs to understand how much a certain driver of biodiversity loss, such as land use or climate change (IPBES, 2019), is caused by one unit of the studied activity. For example, how much land use the purchases of IT equipment causes or how much greenhouse gases are emitted due to business travel. Different approaches, such as the life cycle assessment (LCA) and environmentally extended input output assessment (EEIOA) analysis are presented in Supplementary Information SI1.

**STEP 3: Assess the biodiversity impact caused by the drivers of biodiversity loss**

By further integrating LCA and EEIOA with LCIA methods, such as LC-IMPACT (Verones et al., 2020) or ReCiPe (Huijbregts et al., 2017), the quantity of the drivers of biodiversity loss can be converted to biodiversity loss. To calculate biodiversity impact factors for each driver and activity (in this case the different products and services found from the LCA and EEIOA databases) three steps have to be taken: location analysis of each driver and activity (Step 3a), spatially explicit biodiversity impact assessment in each location of each driver of each activity (Step 3b), and compilation of the biodiversity impact factors of consumption (Step 3c).

**Step 3a: Assess the location of the drivers of biodiversity loss**

Since biodiversity is different in different parts of the world and is affected differently by different drivers of biodiversity loss, the location of each of the drivers causing biodiversity impacts must be understood. This analysis must be conducted separately for each of the activities of the organization in question. An activity might include, for example resource use, heat energy consumption or air travel of the organization. The location of the impact can be analysed with regionalized LCA databases, such as ecoinvent (Wernet et al., 2016) and the World Food LCA database (Nemecek et al., 2019), and multiregional EEIOA databases, such as EXIOBASE (Stadler et al., 2018), EORA (Lenzen et al., 2013) or FABIO (Bruckner et al., 2019). Furthermore, locations of impacts can also be analysed directly with information about, for example, the country of acquisition or by analysing national statistics on the origin of different goods and services.

In this study we used EXIOBASE (Stadler et al., 2021) and ecoinvent (ecoinvent, 2016) to analyse the location of the drivers of biodiversity loss. With EXIOBASE we utilized the open-source tool Pymrio (Stadler, 2021) and with ecoinvent the regionalized assessment method in openLCA software. The detailed process for the location analysis has been described in Supplementary Information SI1.

**Step 3b: Assess the biodiversity impact of the drivers in each location**

When the location of the drivers of biodiversity loss is known, the biodiversity impact factors for the country-specific activities can be calculated. The biodiversity impact factors ($BD_{factor}$) for each driver of biodiversity loss in each impact region $i$ (the region where the biodiversity impact takes place), driven by use in each region $j$ (the region where the activity is used) and each activity $k$, can be calculated by multiplying the location matrix of the drivers of biodiversity loss ($DR_{factor,i,j,k}$, see Supplementary Information SI1) with the biodiversity impact factors for each driver of biodiversity loss ($BD_{lc\text{-}impact}$) for each impact region $i$ from LC-IMPACT (Verones, 2021; Verones et al., 2020) or other similar databases:

$$BD_{factor,i,j,k} = DR_{factor,i,j,k} \times BD_{lc\text{-}impact,i}$$

Information on the harmonization of the categorization of the drivers of biodiversity loss between EXIOBASE and LC-IMPACT is provided in SI Table S2. In terms of the biodiversity impacts of climate change, carbon dioxide, methane, fossil methane and nitrous oxide are included. We chose impact factors that take all climate effects into account for a period of 100 years for both terrestrial and aquatic ecosystems (Verones et al., 2020). With the spatial component missing from the climate change biodiversity impact analysis, we then multiplied the biodiversity impact factor of carbon dioxide with the amount of carbon dioxide equivalent emissions emitted by each activity derived from EXIOBASE. Then we summed the results to derive a total biodiversity impact factor of climate change separately for terrestrial and aquatic ecosystems.

**Step 3c: Compile the biodiversity impact factors**

Total biodiversity impact factors ($BD_{factor,\ total}$) for each region *j* and activity *k* can be derived by summing up the biodiversity impact factors of each impact region *i* and activity *k* in a given region *j:*

$$BD_{factor,total,j,k} = \sum_{i=1}^{n} BDe_{i,j,k}$$

For clarity, we provide an illustration of the matrix $BDe_{i,j,k}$ and the resulting $BDe_{factor,total,j,k}$ in SI Table S3. Now the biodiversity impact factors (biodiversity impact per unit for each activity) are known and can be used to assess the biodiversity footprint after further harmonization of the activity data. The global biodiversity footprint factors we have calculated are provided for further research and applications at the BIOVALENT database (https://oscsolutions.cc.jyu.fi/wisdom/biovalent/) and in the Zenodo repository: https://doi.org/10.5281/zenodo.8369650.

**STEP 4: Harmonize organizational activity data with biodiversity impact factors**

The categorization of the organizational activity data is usually not directly compatible with the LCA classification or the EEIOA activity categorization, which is why the categorizations must be harmonized. Determining a suitable match from the LCA and EEIOA categorization for all financial accounts and activities of the organization can be onerous but it helps when the chosen databases have high activity detail. The harmonization of financial accounts can be done based on the chart of accounts containing information about all accounts in the general ledger of the organization. Further adjustments needed are discussed in Supplementary Information SI1.

**STEP 5: Calculate the biodiversity footprints**

**Step 5a: Calculate ecosystem-specific biodiversity footprints**

The biodiversity footprint is first calculated for each driver of biodiversity loss individually by multiplying the activity data of the organization with the activity specific biodiversity impact factor (biodiversity impact per unit  for each activity) derived from Step 3, and then by summing the biodiversity footprint across the activities within each of the three impacted ecosystem types: terrestrial, freshwater and marine ecosystems. At this stage, it is good to note that biodiversity footprints shall not be directly summed across different ecosystem types, as they have a different number of species, and that is the subject of step 5b.

**Step 5b: Calculate the biodiversity equivalent**

Finally, to arrive at a single biodiversity footprint value for the organization i.e. the biodiversity equivalent, the biodiversity footprints in different ecosystem types can be merged by taking a number of species-weighted average of biodiversity footprints over ecosystem types. As weights we used the estimated share of all plant and animal species that exist in each ecosystem type (Román-Palacios et al., 2022). The biodiversity equivalent (BDe) can then be calculated with the equation:

$$BDe = BF_{terrestrial} \times 0.801 + BF_{freshwater} \times 0.096 + BF_{marine} \times 0.102$$

For example, $BF_{terrestrial}$ is the biodiversity footprint in terrestrial ecosystems, derived from Step5a, and 0.801 is the estimated share of animal and plant species of the world that belong to terrestrial ecosystems. As the biodiversity equivalent estimates the globally potentially disappeared fraction of species caused by the focal organizations' activities, the numbers tend to be very small. To ease the communication of results, they can be presented with the International System of Units metric prefixes such as nano ($10^{-9}$), pico ($10^{-12}$) or femto ($10^{-15}$) biodiversity equivalents i.e. nBDe, pBDe or fBDe respectively. The logic here is similar to the conversion of kilogrammes of carbon dioxide equivalents to tonnes of carbon dioxide equivalents ($tCO_2e$) for easier presentation of results in the context of carbon footprint assessment.

**STEP 6: Assemble the value-transforming financial-environmental impact statement**

Assessing and reporting biodiversity and carbon footprints is a necessary, yet inadequate, step in transforming the operations of organizations (Bracci & Maran, 2013; Maas et al., 2016; Tregidga & Laine, 2021; Veldman & Jansson, 2020). To overcome this challenge, we suggest integration of environmental and financial accounting in organizations to the extent that the environmental accounts transform the actual value of the financial accounts.

In financial accounting, the relevant information is generally reported in an income statement and a balance sheet. For biodiversity and carbon footprint analysis it is the income statement which contains most of the information needed, that is, the incomes and costs of the organization. To transform the value of the financial accounts of the organization, the footprints need to be where the financial income and expenses are located. The balance sheet, which contains information about the organization's assets, could be used in natural capital (Houdet et al., 2020) and handprint (Pajula et al., 2021) accounting, but these fall outside the scope of our current paper. However, the BIOVALENT method could also be used to understand the assets of the organization, for example by reporting the land assets of the organization and reporting any changes in the biodiversity equivalent values of the land assets in the balance sheet. To transform the financial value, the biodiversity and carbon footprints need to have a cost that becomes visible in the income statement. One way of putting a price on the biodiversity and carbon footprints is to finance offsets matching the footprints. We discuss the debate around the integration of financial and environmental accounts more extensively in Supplementary Information SI1.

## 3. Results: The biodiversity and carbon footprint of the University of Jyväskylä

To demonstrate the use of the Biodiversity Equivalent Impact Assessment method BIOVALENT, we conducted an example study. We assessed the biodiversity and carbon footprint of the University of Jyväskylä in Finland. The university has 14 600 degree students, 2 800 staff members and 230 million euros in annual turnover (University of Jyväskylä, n.d.). The university's size is comparable to a large corporation and as such provides an interesting and suitable platform for testing the utility of the BIOVALENT method.

We utilized financial accounts in addition to other activity data for energy use and business travel and the BIOVALENT methodology to assess the carbon and biodiversity footprints of the University. To illustrate the results, we aggregated the activity data from all activities to 16 broad categories and calculated the relative contribution of each category to the carbon and biodiversity footprints. The total annual biodiversity footprint remained nearly the same in 2023 when compared to the total biodiversity footprint in 2019 (58 nBDe). In turn the total annual carbon footprint decreased by 15 % from 22 723 $tCO_2e$ in 2019 to 19 405 $tCO_2e$ in 2023. Even though heat and electricity consumption carbon and biodiversity footprints decreased, the use of biomass in energy production paired with the increase of purchases, for example in terms of IT supplies and food, kept the total biodiversity footprint stable. (Figure 4).

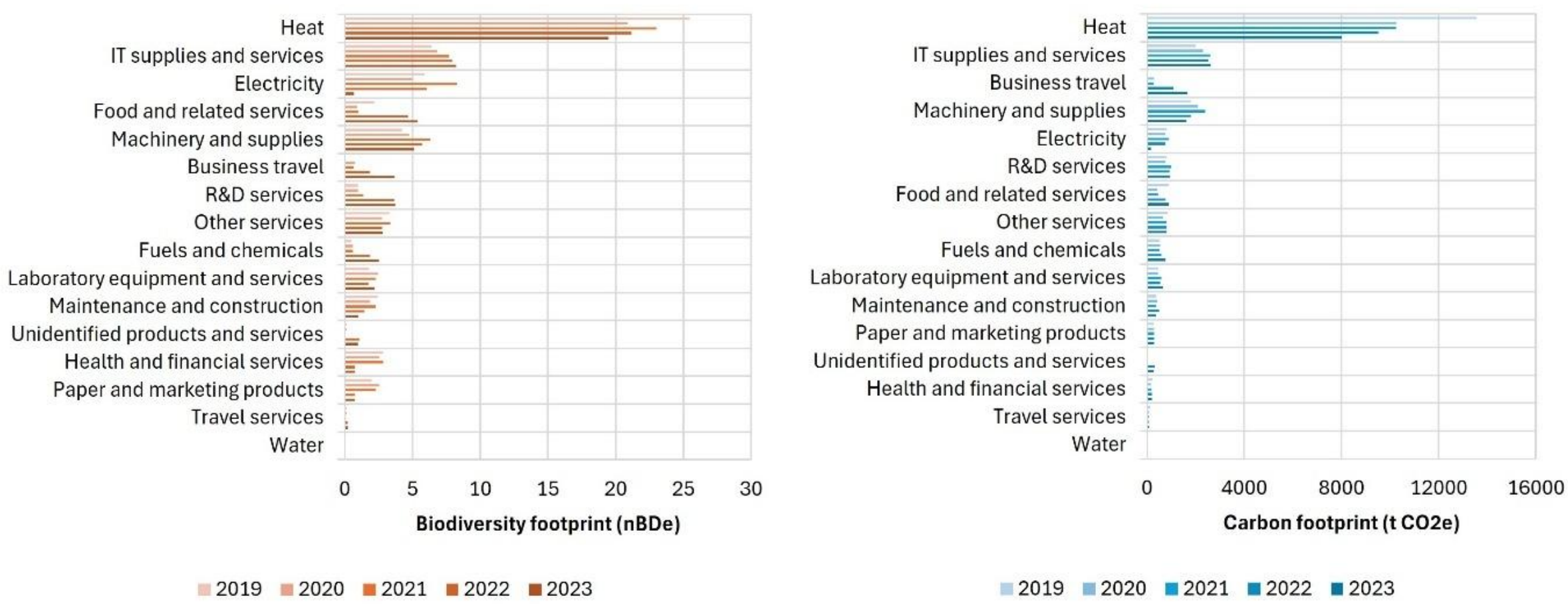

**Fig. 4** The contribution of 16 consumption categories to the annual biodiversity and carbon footprints of the University of Jyväskylä during 2019-2023. The data for Figure 4 can be found in Supplementary Information SI2.

To illustrate Step 6 of the methodology we assembled a value-transforming financial-environmental impact statement for the university in 2023 (Table 1). To evaluate the offsetting cost of the carbon footprint, we used the World Bank's carbon pricing statistics for the European Union, which was around 96 US\$/t$CO_2$e in 2023 (World Bank, n.d.). As no such statistics are available for biodiversity footprints, we developed one to demonstrate the idea.

As stated above, a desirable characteristic of the biodiversity equivalent is that it provides a unified way of measuring biodiversity loss across the planet. While we first used the biodiversity equivalent to measure biodiversity loss due to the activities of the University of Jyväskylä through drivers like continued land use, here we reverse the logic and use the same land use biodiversity impact factors to estimate biodiversity gain achieved if the continuous exploitation is ceased for the purpose of offsetting biodiversity loss. Biodiversity offsetting is complicated business (Moilanen & Kotiaho 2018, 2021) and the suggestion made here about global offsetting is very controversial. Please note that we purposefully illustrate the point of value-transforming accounting here with the overly simplified assumptions and will provide a rigorous scrutiny of the suggested global offsetting in the light of the fifteen operational considerations of biodiversity offsetting elsewhere (Kalliolevo et al., 2025).

For making the point we used the LC-IMPACT database (Verones et al., 2020) to calculate the area of land used for intensive forestry that should be permanently removed from use (i.e. protected) in Finland or in Brazil to offset the global biodiversity footprint of the University of Jyväskylä in 2023. We use Finland and Brazil as examples to demonstrate how, from a global species richness perspective, the biodiversity equivalent operates at two very different geographies in value-transformation of the accounts. Although, the biodiversity footprint of the University is spread out across the world with different amounts of impacts in different countries, the characteristics of the biodiversity equivalent allow their offsetting anywhere around the world (Kalliolevo et al., 2025). The detailed information for assessing the biodiversity offsetting values for the specific case of University of Jyväskylä is provided in the Supplementary Information SI1. The results show that if the cost of offsetting is distributed across 30 years similar to the depreciation of large investments, the annual cost for the university would be around 435.78 M€ if the offset was completed in Finland and 0.60 M€ if it was completed in Brazil.

Finally, building on earlier research (Houdet et al., 2020; Nicholls, 2020), we compiled the financial-environmental impact statement. By amending the statement with the biodiversity and carbon offsetting values, we arrived at the value-transforming integration of financial and environmental accounts (Table 1). In financial accounts, net income is generally the deduction of costs from revenue. By adopting the same logic, the net biodiversity and carbon footprint is the deduction of the footprints from their respective offsets. The integrated financial-environmental impact statement can be used to quickly deduce the economic and environmental position of the organization and their interlinkages through the valuation of footprints.

**Table 1: The financial-environmental impact statement of the University of Jyväskylä in 2023. As units we use thousands of euros (k€), tonnes of carbon dioxide equivalents ($tCO_2e$) and nano biodiversity equivalents (nBDe).**

| | Financial footprint (k€) | Carbon footprint ($tCO_2e$) | Biodiversity footprint (nBDe) |
|---|---|---|---|
| **Revenue** | | | |
| **Government funding** | 152 151 | - | - |
| **Other revenue from operations** | 80 720 | - | - |
| | | | |
| **Costs / Footprints** | | | |
| **Staff costs** | 166 856 | 46 | 0.15 |
| **Depreciation** | 2 503 | 763 | 2.37 |
| **Grants** | 3 854 | 191 | 0.59 |
| **Raw materials, equipment, and goods** | 10 031 | 3088 | 11.15 |
| **Services** | 15 012 | 2962 | 11.35 |
| **Rents** | 28 743 | 158 | 0.35 |
| **Travel** | 6 408 | 1992 | 5.28 |
| **Other** | 10 335 | 10223 | 26.33 |
| **Total Costs / Footprints** | 243 742 | 19423 | 57.58 |
| | | | |
| **Losses and Gains** | | | |
| **Fundraising** | 227 | - | - |
| **Investment gains and losses** | 2 229 | - | - |
| **Appropriation** | -70 | - | - |
| | | | |
| **Impact pricing** | | | |
| **Carbon offsets** | 1802 | -19 423 | - |
| **Biodiversity offsets in Finland** | 364762 | - | -57.58 |
| **Biodiversity offsets in Brazil** | 506 | - | -57.58 |
| | | | |
| **Net Income / Footprint** | | | |
| **Net footprint without offsets** | -8486 | 19423 | 57.58 |
| **Net footprint with offsets in Finland** | -373 248 | 0 | 0 |
| **Net footprint with offsets in Brazil** | -8 992 | 0 | 0 |

## 4. Discussion

We believe that the scattered field of biodiversity indicators and the complexity in how the results of biodiversity footprinting are communicated are key challenges in the uptake of biodiversity footprinting in organizations. Furthermore, there is a lack of global consensus and direction, even though some initiatives are aiming to standardize the measurement of biodiversity impacts (Nature Positive Initiative, 2025; UNEP-WCMC et al., 2022). A vital obstacle to date seems to be the general reluctancy to try to simplify the very nuanced nature of biodiversity such as genetic, functional, species and ecosystems. The idea of the biodiversity equivalent that we presented in this perspective may appear controversial, yet in our understanding it is a much-needed opening of discussion for a unified global indicator that could be used in high-level decision-making. The BIOVALENT methodology we have presented here together with the biodiversity equivalent can be applied in all kinds of organizations around the world. To help the adaptation we provide access to global biodiversity impact factors (https://oscsolutions.cc.jyu.fi/wisdom/biovalent/).

At the same time, like most other indicators, the biodiversity equivalent does not capture all variability among living organisms, including the genetic, species and ecosystem diversity (Díaz et al., 2015). The biodiversity equivalent could also drive potentially controversial actions from the viewpoint of regional biodiversity. The example we presented to offset the biodiversity footprint of the University, suggests that offsetting the global biodiversity footprint could be easier and cheaper in a country with high species richness (Brazil) than it is compared to a country with low species richness (Finland). The problem is similar to the issues presented by the global carbon offsetting market (Rizzi & Krause, 2025). Researchers have also shown that 25 carbon offsetting schemes have not been able to achieve additionality (Becken & Mackey, 2017;
26 Cames et al., 2016), a key issue also in biodiversity offsetting (Moilanen & Kotiaho, 2018).
27 Careful consideration is needed if a biodiversity equivalent indicator were to be used globally for
28 offsetting activities. While we understand the potentially dangerous implications of the
29 biodiversity equivalent, we encourage debate on how a more global viewpoint on biodiversity
30 might help humanity in driving change to stop global biodiversity loss. We believe that the
31 BIOVALENT method encourages a broader perspective to understand the biodiversity impacts
32 of supply chains and thus could help mitigate the unsustainable appropriation of resources from
33 the Global South to the Global North (Hickel et al., 2022).
34
35 Another interesting problem that needs to be addressed by the complexity reduction of the
36 biodiversity equivalent is how it reacts to offsetting efforts in different drivers of biodiversity
37 loss. While it can be beneficial for unified decision-making to combine the different drivers of
38 biodiversity loss, such as land use and climate change, under a single indicator, one also needs to
39 understand that these phenomena affect biodiversity in different ways. For example, we can
40 consider that the impacts of land use change do not cumulate over time, if no further land is
41 taken into use. However, this is not the case with climate change. As long as emissions to the
42 atmosphere exceed the capacity of the sinks to bind emissions, the concentration of greenhouse
43 gases will increase, creating further global warming, and thus posing increased risks to species
44 (Urban, 2024). Another example is the temporal aspect: the impact of land use change might be
45 realized almost immediately, while the impacts of climate change might be realized in the course
46 of a century. Thus, in addition to the biodiversity equivalent, focus should also be given to the underlying results provided by the drivers of biodiversity loss (Bromwich et al., 2025; Martínez-Ramón, 2024), which might be subject to different kind of mitigation and offsetting decision.

An interesting prospect is how the biodiversity equivalent might be used by organizations to analyze their contributions to the global goal of stopping biodiversity loss, similar to carbon footprint analyses being compared against the Paris Agreement's 1.5-degree target. One of the main targets in the Kunming-Montreal Global Biodiversity Framework states that "by 2050, the extinction rate and risk of all species are reduced tenfold" (CBD, 2022). The biodiversity equivalent, which essentially estimates the extinction risk of species, could be utilized to understand whether a company's activities bring us closer or further away from this goal. Indeed, Rounsevell et al. (2020) already suggested that a biodiversity target based on species extinctions might provide much-needed unity to design effective policies for stopping global biodiversity loss.

From an accounting perspective, the utility of BIOVALENT and other biodiversity footprint methodologies need to be scrutinized whether they serve the requirements set in regulatory and voluntary frameworks such as the EU Corporate Sustainability Reporting Directive (CSRD), Taskforce on Nature Related Financial Disclosures (TNFD) and Science-based Targets for Nature (SBTN). Overall, it seems that biodiversity footprinting is a practical tool that can be used to comply with the general requirements of such frameworks (Kundi et al., 2026). Furthermore, even though we argue that decision-making needs a value-transforming integration of financial and environmental accounting, it has been argued that monetizing nature, for example in the form of offsets, could legitimize further destruction of nature (Redford & Adams, 2009; Spash, 2015). In addition, Nedopil (2022) argued that we need regulatory limits to exploitation of nature since “finance only values limited aspects of nature”.  On the other hand, the approach here aims to set a value to the biodiversity impact caused by an organization, rather than valuing nature per se, which might work towards mitigating biodiversity impacts, rather than making nature a property of the economic system. Nevertheless, a broad array of actions is needed to incorporate the diverse values of nature, not only financial value, into decision-making (Vatn et al., 2024).

As Booth et al. (2024) suggest, transformative changes in organizations can take place when action is applied at different scales from corporate-specific to sectoral and value chain actions. We provide an example for organizations on how to give financial value to environmental impacts and consequently communicate the importance of the results derived from the environmental accounts. Methodologies such as BIOVALENT provide opportunities for companies to provide quantified information for their nature transition plans required for example in the EU CSRD regulation, aiming to understand how their business models create biodiversity impacts, and how in turn the business models are affected by nature-related impacts, dependencies, risks and opportunities (EFRAG, 2025). We also provide valuable insights for policymakers on how mandatory offsetting and taxation of environmental impacts might drive changes in the financial positioning of organizations and consequently drive transformative changes in them. Indeed, the integration of environmental and financial accounting is not only a technical accounting issue; it is also a public policy issue (Nicholls, 2020). Previously, it has been stressed that value-transforming economic instruments to protect biodiversity, including biodiversity offset programs, do not and most likely cannot operate without robust regulation and government involvement (Boisvert, 2015; Koh et al., 2017, 2019; Kujala et al., 2022; Vatn, 2015).

Finally, we open the debate by arguing that as the biodiversity equivalent provides a unified way of measuring biodiversity loss across the planet, it may also provide a location-independent unified way for offsetting the loss. We argue that the biodiversity equivalent measures biodiversity loss potential similarly to how the location independent carbon dioxide equivalent measures the global warming potential (see Supplementary Information SI1). Merely focusing on local biodiversity offsetting might not be enough to rapidly halt biodiversity loss due to the globalized nature of the biodiversity impacts of consumption (Balmford et al., 2025; Lenzen et al., 2012; Marquardt et al., 2021). We think that extensive adoption of value-transforming integration of financial and environmental accounting is essential to influence decision-making in organizations and to facilitate the much-needed transformative change in our production and consumption practices in support of planetary well-being (Díaz et al., 2019; Kortetmäki et al., 2021).

**ACKNOWLEDGMENTS**

We thank the Strategic Research Council at the Academy of Finland (JY 364448), Green Carbon Finland Ltd, The Finnish Innovation Fund Sitra and SOK Corporation for funding the development of the methodologies. We thank the administration at the University of Jyväskylä for giving access to the consumption data, the Biodiversity Footprint Team members as well as Matti Toivonen, Janne Peljo and Hanna-Leena Pesonen for feedback.

The authors declare no competing interests.

## SUPPLEMENTARY INFORMATION

The supplementary information provides supplementary methods detailing the steps of the development of the BIOVALENT database as well as additional details to calculate the results presented in the main manuscript. The supplementary information also provides the data behind figure 4 of the main article and figure S1 of the Supplementary Information S1.

**Supplementary Information**

Supplementary information is linked to this article on the *JIE* website:

**Supplementary Information S1:** This supplementary information provides supplementary methods detailing the steps of the development of the BIOVALENT database as well as additional details to calculate the results presented in the main manuscript. Supplementary discussion points are also included.

**Supplementary Information S2:** The supplementary information provides supplementary data on the case study results of the main manuscript, namely figure 4 and figure S1. The full BIOVALENT can be accessed in a database format in

https://oscsolutions.cc.jyu.fi/wisdom/biovalent/, and as full data files in Zenodo at: https://doi.org/10.5281/zenodo.8369650

477
478
479
480